# AN APPROACH ON THE MODELLING OF LONG ECONOMIC CYCLES IN THE CONTEXT OF SUSTAINABLE DEVELOPMENT

## Cristina TĂNĂSESCU [1], Amelia BUCUR [2], Camelia OPREAN-STAN [3]

*[1,2,3] Lucian Blaga University of Sibiu, Romania*

**Abstract**

*One of the themes that have been approached more and more within the specialised literature is being represented by economic cycles. The analysis of these is very useful in the long term predictions, in finding solutions for the economic raise and for detecting the economic crisis. At the same time, it is underlined in a lot of scientific and research papers, the importance of the sustainable development in the present and future society. In this paper we intend to bring contributions to the study of the cycles of a sustainable economy and we will analyse it having in mind the purpose of creating the sustainable economy. We will demonstrate the fact that curves that represent graphically all these, are not simple logistics anymore, bi-logistics or multi-logistics curves, but curves in plan that are obtained by composing logistics functions with the function of the sustainable development or with the function that shapes the economic component of it mathematically. We will present an interpretation of mathematic models within the frame of the sustainable development.*

## 1. Introduction

It is said that we have been living in a continuing error for centuries. Whether we considered the environment as a supernaturally given, an unknown, a flood that hit us, we do not know to stay away or we did not take all into account, incapacitating it, humiliating it, hurting it and deteriorated it, by us, the people, whenever we had the opportunity. Reflecting a lot in such situations, Horace wrote in "Epistulae":"*Naturam expelees furca, taman usqve recurret*"  (banishing nature with the fork, but she returns to race), and Ovidiu show itself, *"Gutta cavet stoned non vi, sed saepe cadendo"* (drop penetrates stone not by force but by continuing to drop). 17 centuries later, the great Francis Bacon, in "Novum Organum", understood the core problem by offering a solution which, by its depth and intelligence, overpassed well beyond the limits of his time: "Nature, to be commanded, must be obeyed."

*[1] Faculty of Economic Sciences, cristina.tanasescu@ulbsibiu.ro*

*[2] Faculty of Sciences/Department of Mathematics, amelia.bucur@ulbsibiu.ro*

*[3] Faculty of Economic Sciences, camellia.oprean@ulbsibiu.ro*

A deep philosophy, unfortunately unnoticed even by his contemporaries, and followers ... It took the great energy crisis and resource in `70s in Europe and the world to become aware of such a state of affairs. It took the Meadows with the paper "**Limits of growth**", the brilliant American professor of Romanian origin Nicolas Georgescu-Roegen, in his great book "entropy and Economic Process" written in the 1970 years for the XXI century, to realize that we are on a fundamentally wrong way, privileging only individual profit, but marginalizing that individual profit and the integrating public component by "internalising externalities". However, it is sustainable development which requires a new paradigm of profit, another conceptual system of values, is that which can help us. And globalization, discussed in such forward-looking perspective, takes place especially, "inside to outside" and not particularly „outside to inside"as if caused solely by the very strict interests of globalization, overly-off of an equity approach only from the organization.

Although scientists have long warned of the consequences of worsening brought by  pollution on the planet, the decisions to prevent this are being postponed or considered unimportant. Lately, however, people have begun to realize that the environment is in a continual degradation. People began to see how complex the concept of sustainable development is, which is not only an environmental issue but it is also a matter of technology, economics, international trade, human rights, health etc.

## 2.  Economic cycles – between theory and possible configurations

The economic life evolves cyclically; man prints his own archetype, his DNA.

***By continuous cycles of the economic process we understand the intrinsic property that allows it to present the values of its mood variables,***

*that modifies with a certain frequency, within the frame of the dynamic trend report, and more rarely, within the frame of report of quantitative value. Its main qualitative determinations are the following:*

- Alternation: the economic process submitted to the continuous repeated cycles supports, alternatively, processes of economic increase and decrease;
- Periodicity: the economic process submitted to the cyclicity has some aproximately returns, under the quality aspect, to some previous values, after a period that could be estimated. One have to mention that the frequency must not be estimated within a strictly calendaristic frame, but within the frame of a specific economic time, that depends on the evolution of various factors regarding the economic continuous cycles.
- Inherent process: the economic process will always overlap to this cyclic nature and will not be able to take place outside the phenomenon.
- Cumulation: the economic process can alternate, within the frame of the phenomenon of cyclicity, on the basis of a cumulative process. During this cumulative process, certain features and disorders reach their limits, in time. By reaching these limits the economic process passes to an alternative tendency.
- Self-regulation: the cyclicity is a cibernetic economic process and it is characterised by self-regulation. The alternating phases of the economic process is due exactly to the fact that once the economic process has entered a phase, all the influential factors start acting with the purpose of getting out this phase. The exit will take place when the phase has reached the cumulative limits.

The cyclicity is therefore an immanent feature of the economic process.

### 2.1. Kondratieff Cycle and century-old trend

In a large rolling over the way in economic development, when all the foundations of economic life - present and past - are put back into question, and all the lessons of history do not help those who have learned them, the precise understanding of long cycles, which seem to obey certain laws or tendentious rules, is an opportunity for learning about the direction in which the contemporary world economy is moving. General development of the economy (based on data relating to advanced countries) shows that the country's productive forces takes place in the form of "waves" (cycles) long by 40-60 years. This period is dominated by a technical way of production of certain technology. Some experts identify the big "waves" with the history time of the outbreak, maturation and exhaustion values of an industrial revolution.

In the development of each economy - and this especially in light cycles – two main phases are distinguished: an ascending phase and a descending phase, each with a duration of 20-30 years.

In the ascending phase the implementation and technical operation of the new way of production occurs, business activities - based on new technologies - take place at this stage, with increased efficiency. Within the period of 20-30 years there are being observed: the dominance of years of prosperity, relatively high rates of economic growth, rising living standards, high employment level, etc.. The period of transition from the old way to the new production techniques is marked by a structural crisis, a period which is extended during downward phase. In this phase of long economic cycles, initially, some showing signs of exhaustion values of favorable growth factors occur. There is a tendency of decreased efficiency and rate of profit. It is the sustained period of scientific research to find solutions to streamline the production process. It also marks a transition to new production techniques.

If we look at the parallel fluctuations of investment and scientific research over the two stagesof the long cycle, we will note the following:

- The *ascending phase* entails a sustained and effective investment process based on previous scientific discoveries (in the downward phase). In this phase of the research recorded a rebound (in intensity) and a high efficiency of investments made on the basis of previous findings.

- The *downward phase*, produces instead a relative decline in investment efficiency (of production) and an increase in the scientific and technological research. Now the structural crisis manifests (technology, industry, etc..), crisis which is specific for transition from one technology to another, from one stage of technical progress and scientific and technical office to another. Available statistics seem to show that the peaks of scientific discoveries and technological innovations have placed in downward phases of long cycles.

Innovation is the exogenous factor of production "that makes economic life to be cyclical in nature". Innovation should not be confused with invention. The invention of a new manufacturing process becomes innovation as long as it is not brought into production. Innovation is a process of industrial transformation that continuously revolutionizes the economic structure from inside, destroying continuous its old items and creating new items continuously.

A short term decline may be experienced during a long-term expansion, as may be a small increase within long-term contracts. These large oscillations are due to the specific dynamics of various components of social and economic system (population, employment, fixed assets, raw materials and energy resources, economic mechanism), the interaction between them, the contradictions that have appeared in Economic and Social Development , the delayed reaction of economic agents in different modification, the inertia that is shown by some components.

The main tendency within the organization of firms will be an integrated approach linked to the introduction of new clean technologies and personnel training, or business organization having as a fundamental principle combining technology with the organizational redevelopment. Regarding the company, cross-technology can do exactly that difference that is needed between products or services and those offered by competitors. In addition, as far as the use of new technologies entail welfare increases, their non-utility generates penalty. So, in terms of how organizations must adapt to new technologies, consider: $AO + IT = IVO$, an organization that is adaptable and interactive cross-technology lead to an innovative organization and long-term value, while $OO + NT = EOO$, which means the old organization and the new eco-technologies lead to an expensive old organization.

## 2.2. Economic cycle and sustainable development approach

Although sustainable development was initially meant to be a solution to ecological crisis caused by intense industrial exploitation of resources and continued environmental degradation and primarily looks to preserve environmental quality, today it expanded the concept of quality of life in its complexity and the economical and social aspect. The purpose of sustainable development is here and now, the concern for justice and equity between states, not only between generations.

The concept of sustainable development starts from the idea that human activities are addicted to the environment and the planet's resources. Health, social security and economic stability of society are the major elements in the definition of quality of life.

Sustainable development aims and tries to find a stable theoretical framework for taking decision in any situation, in which can be found a report like person/environment, be it environmental, economic or social (Brown, 2001).

The international community decided to solve environmental problems through global collective action, which sought to define and implement them through an appropriate international framework. This framework for action at the international level has shaped over time and it is in a dynamic evolution, including legal measures:

- binding as treaties or conventions;
- non-binding, in the form of declarations, resolutions or sets of guidelines and policy guidelines, institutional arrangements and mechanisms for sustainable financing

Sustainable development is a pretty hard concept to be define. Because of its continuous evolution it is twice harder to define. Usually when we talk about sustainable development we take into consideration three components: economy, society and environment.

It is crucial to emphasize the fact that sustainable development isn't just about the environment. The social part of sustainable development is increasingly becoming very important for governments, organizations and civil society and it cannot be omitted from any serious discussion about sustainability. An important place on the sustainable development agenda holds the transformation of economy which should help to accomplish the sustainability. Sustainable development is a horizontal project which requires an integrated decision.

These three elements of sustainability introduce new approaches on sustainable development which will create opportunities for work places in new domains like social and environmental and for the recovery of the differences between east and west, on the equilibrium principle.

As a consequence of the population growth and the consumption of natural resources under the conditions in which there is a permanent decrease of the un-renewable resources of the planet, the human kind searched for new modalities of development.

So it reached a development pattern that aims a balance between economic growth, life quality and environment protection on medium and long term, without increasing the consumption of natural resources beyond Terra`s capacities. This is sustainable development.

Sustainability is a paradigm in which future is seen as a balance between economic growth and environment protection and, on this basis, satisfying social development requests, not only present requests but also future requests with the aim of developing and improving life quality.

These challenges request a change of the industrial policy directed towards society and environment. The business environment should play an important role in order to respond to these necessities and the social and environmental part is in the middle of the proceedings regarding the common social responsibility.

The positive progress recorded in the European Union industrial production as well as the decrease of some pollutant emissions prove, once more that, high competitiveness and environmental protection can be obtained by industry with the support of an adequate mix of policies.

Society needs to create new business opportunities. Thus, the focus on environmental protection has allowed European companies to be the forefront in environmental technologies and encourage them to develop sustainable production that is based on life cycle analysis. Proximity to consumer needs, especially food quality and safety requirements, have created new market niches.

Strong economic growth provides resources to achieve social and increasing environmental needs. Sustainable organizations must be characterized by the fact that they practice a performance management that retains biodiversity and that are profitable in the long term. On the human dimension, particular attention should be paid to the marketing systems, which have a strong influence in encouraging or discouraging the adoption of sustainable practices.

It is very important to find solutions to solve delays in the three dimensions of sustainable development: economic issues must be balanced with social welfare issues and the environment. So it is a necessary exercise to promote nature conservation and poverty eradication.

## 3. A mathematical model of sustainable development. Function of sustainable development

We measure Sustainable Development as :

$$f(t) = \left(f_1(x_1(t), \ldots, x_n(t)), f_2\left(x_{n+1}(t), \ldots, x_{n+m}(t)\right), f_3(x_{n+m+1}(t), \ldots, x_{n+m+p}(t))\right)$$

$$= \left(\frac{\sum_{k=1}^{n}[\pm p_k(t)x_k(t)]10^{i_k}}{n}, \frac{\sum_{k=1}^{m}\left[\pm r_k(t)x_{n+k}\right]10^{j_k}}{m}, \frac{\sum_{k=1}^{p}\left[\pm q_k(t)\,x_{n+m+k}(t)\right]10^{l_k}}{p}\right),$$

(1)

$$f_t = (f_{1t}, f_{2t}, f_{3t}); \quad f^{'}(t) = f_t^{'} = \frac{f_{1t}+f_{2t}+f_{3t}}{3},$$

(2)

$$f: N_{2010} \to [0, \infty) \times [0, \infty) \times [0, \infty)$$
$$N_{2010} = \{2010, 2011, 2012, \ldots\}$$
$$f^{'}: N_{2010} \to [0, \infty)$$

where: t is time, expressed in years.
We take as baseline the year 2010, because non-renewable resource reserves are being estimated at the time of the year.

The expression of this feature vector includes the components:

$f_1$ – economical function,

(3)

$f_2$ – social function,

(4)

$f_3$ – environment function

(5)

Indicators can be (see [2]):

$x_1$ - indicator of global settlement (6)
$x_2$ – indicator of quick ratio (7)
$x_3$ – profitability of the capital employed,

(8)

$x_4$ - gross margin sales

(9)

$x_5$ – indicator of economic profitability

(10)

$x_6$ – indicator of efficiency of production costs

(11)

$x_7$ – indicator of the share of wages in costs

(12)

$x_8$ – indicator of efficiency rate of the total costs

(13)

$x_9$ – efficiency of using production capacity

(14)

Indicators $x_i, \quad i = \overline{n+1, n+m}$ can have the following meanings:

$$x_{n+1}: N_{2010} \to \Omega_{n+1} \subseteq [0, \infty)$$ 

(15)

$x_{n+1}(t)$ - expenses with insurance and social protection (CAS, unemployment, CASS, Occupational diseases)

$$x_{n+2}: N_{2010} \to \Omega_{n+2} \subseteq [0, \infty)$$

$x_{n+2}(t)$ - expenses for equipment and protective materials

(16)

Indicators $x_i, \quad i = \overline{n+m+1, n+m+p}$ can have the following meanings:

$x_i: N_{2010} \to \Omega_i \subseteq [0, \infty)$ indicator of renewable resources function (wind, solar, hydro, biomass, geothermal)

(17)

$$i = \overline{n+m+1, n+m+5}$$

$x_{n+m+6}: N_{2010} \to \Omega_{n+m+6} \subseteq [0, \infty)$ frequency index of exposure to noxious, 

(18)

$x_{n+m+7}: N_{2010} \to \Omega_{n+m+7} \subseteq [0, \infty)$ is an indicator that reflects the amount of emissions of pollutants into the atmosphere,

(19)

$x_{n+m+8}(t)$ = quantity of emissions (t year), or

quantity of air emissions with destructive potential on the ozone layer (related to year t), or

quantity of air emissions with effect on global warming

(20)

$$x_{n+m+9}: N_{2010} \to \Omega_{n+m+9} \subseteq [0, \infty)$$ is an indicator which reflects

work environment;

$x_{n+m+9}(t)$ = noise and vibration level; quantity of emitted radiations; quantity of heat and light. (21)

We extend the above model by using interpolation functions on the interval [2010, 2010+10 z] where z is a natural number bigger than 1

Let *h : [a,b]→ R* and let *t₁, t₂,..., tₙ₊₁* (n+1) distinct points in range *[a,b]*, called nodes. Problem interpolation function h in nodes *tᵢ* $i \in \{1, ..., n+1\}$, consists in determine a function g *: [a,b]→ R,* a class of known functions, with the property $g(t_i) = h(t_i) \ i \in \{1, ..., n+1\}$.

The most used class of interpolation functions is the polynomials class, due to the ease with which derives and integrates.

**Theorem 1** *Let h: [a,b]→ R and i t₁, t₂,..., tₙ₊₁ (n+1) nodes in range [a,b]. Then there is an unique polynomial* $P_n$, *of degree n, that integrates function h in nodes tᵢ,* $i \in \{1, ..., n+1\}$

$$h(t_i) = P_n(t_i), \qquad\qquad i \in \{1, ..., n+1\}.$$

(22)

This polynomial is called the Lagrange polynomial interpolation.
The error in each point is

$$E(h,t) = h(t) - P_n(t).$$

(23)

Obviously $E(h, t_i) = 0, i \in \{1, ..., n+1\}$.

**Theorem 2** *if* $h \in C^{(n+1)}[a, b]$, *then for any* $t \in [a, b], exist\breve{a} \ \gamma_t \in (a, b)$

*So that*

$$E(h,t) = \frac{h^{(n+1)}(\gamma_t)}{(n+1)!} \prod_1^{n+1}(t - t_i).$$

(24)

Let $g : [2010, 2020] \to R$ with the property that $g(t_i) = f'(t_i), i \in \{2010, 2020\}$. $g$ might be an interpolation polynomial, for example the Lagrange polynomial interpolation. We define the function $G : [2010, \infty) \to R$ periodic, with the main period $T=10$, so that:

$$G(t) = g(t - 10k), \ t \in (2010 + 10k, 2020 + 10k], \ \ k \in N.$$
(25)

The function being defined as such constitutes a generalization of the function adjunct to sustainable development function from formula (2).

## 4. A mathematical pattern for the long cycles in the context of sustainable development

In establishing a mathematical model of sustainable economic cycles, we considered that in general, in economy, no sequence of events is repeated at the same level of manifestation, so this is why the probability models are best suited for modeling the economic phenomena of society. Statistical probability models, based on the fact that any causal relationship between events is characterized by a particular property.

We also considered the fact that the economic process is clearly entropic and not mechanical. An accurate evaluation of the scale and dynamical complexity that characterizes the economy doesn`t require the use of fragmentation-promoted method from the position of a mechanistic conception of the natural world - but the use of the sintetico-functional method, capable to provide adequate information about the dynamic interaction of elements of an organized overall of features, in relation to different purpose. We also considered the well known fact that there isn`t linear legitimacy between causes and effects.

Kondratiev cycles that we wish to consider in the context of sustainable development in this paper, are generated by the diffusion of innovations in various fields. This has been thoroughly studied in the field literature. It has been shown that economic growth is modeled by functions that are solutions of the logistic Verhulst equation:

$$\frac{dy}{dt} = ay\,(y - y_0)$$

(26)

where $y = y(t)$ is the production function, $y_0$ is a limit of the market volume, and *a is a constant*.

Marchetti used the logistic approach to analyse the evolution of different types of technologies (for instance the amount of railways and paved roads or the length of telegraph wires), and connected it with Kondratieff long business cycles (Marchetti, 1986). His paper was characterised by Devezas & Corredine (2001) as "the most decisive contribution to the revival of Kondratieff waves".

Equation solution (26) logistic curve equation is (Kucharany, D., Guio R., 2009):

$$y = \frac{y_0}{1 + c \exp\,(-ay_0 t)}$$

(27)

Through a change of the function $F = {}^{y}/_{y_0}$ we get

$$\frac{dF}{dt} = aF(1 - F), \qquad F = \frac{1}{1 + c \exp\,(-at)}$$

(28)

The function values y have the first time a slow growth, then the growth is accelerating to a point of inflection, then tended to slow the pace to a limit that is no longer exceeded (Fig.2).

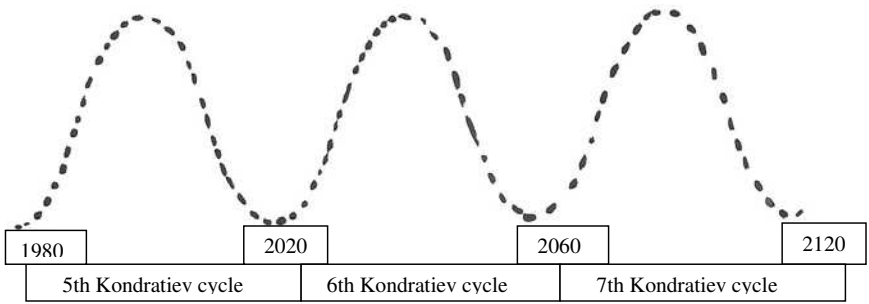

Fig.2

But in terms of sustainable development, the economic cycles will no longer be modeled by logistic functions, bi-or multi-logistic, but the functions of the form.

$$y_{DD}(t) = (y \circ G)_{(t)}$$

(29)

obtained by composition of the relation (27) with the function of the relation (25) of the mathematical model presented in Section III of this article. Curves through the functions   represented will have a different form to those represented in Fig.2, as

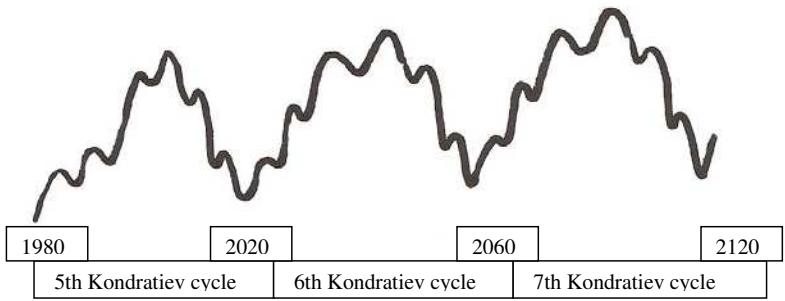

| 1980 | 2020 | 2060 | 2120 |
|------|------|------|------|
| 5th Kondratiev cycle | 6th Kondratiev cycle | 7th Kondratiev cycle | |

Fig.3

Curves will have more points of inflection, points where growth is changing. If we are interested in doubling the time required for production $(t_2)$ we may apply the formula:

$$t_2 = -\ln\left[\frac{k - 2y_0}{2(k - y_0)}\right]\frac{1}{y}$$

(30)

If it is desired to determine the time required to reach a percentage p% of the output limit *k* we may apply to the relationship [ ]

$$t_p = -\ln\left[\frac{k - y_0(1-p)}{p(k - y_0)}\right]\frac{1}{y}$$

(31)

## 5. Conclusions

1. Graphics functions will have intermediate, partial decreasing and increasing phase, towards logistics functions y graphics.

2. Functions are almost periodical [6].

3. G function values can be interpreted as delay or acceleration factors.

Open problems

1. Replacing the G function with other functions of interpolation,which would have values that approximate the real ones with smaller error than the function proposed in the formula (25).

2. Estimate the aggregate diffusion curve with a more or less rapid take-off and a long or short stage of saturation

3. Establish the maximum elements in abstract, competitive economies, within the context of sustainable development.